\documentclass{epl}
\usepackage{amssymb}
\usepackage{amsmath}
\usepackage{epsfig}

\newcommand{\cn}{\rm cn}

\newcommand{\sn}{\rm sn}
\newcommand{\dn}{\rm dn}

\title{Driven Macroscopic Quantum Tunneling of Ultracold Atoms in Engineered Optical Lattices}
\author{ Ramaz Khomeriki\inst{1,2} \and Stefano
Ruffo\inst{1} \and Sandro Wimberger\inst{3}}

\institute{                    
  \inst{1} Dipartimento di Energetica ``S. Stecco'' and CSDC,
Universit\`a di Firenze and INFN, Via S. Marta, 3, 50139 Firenze, Italy \\
  \inst{2} Department of Exact and Natural Sciences, Tbilisi State University, 0128 Tbilisi, Georgia \\
  \inst{3} CNR-INFM and Dipartimento di Fisica ``E. Fermi'', 
Universit\`a degli Studi di Pisa, Largo Pontecorvo 3, 56127 Pisa, Italy
}

\begin{document}

\maketitle

\begin{abstract}
Coherent macroscopic tunneling of a Bose-Einstein condensate between two parts of an optical lattice separated by an energy 
barrier is theoretically investigated. We show that by a pulsewise change of the barrier height, it is possible to switch between 
tunneling regime and a self-trapped state of the condensate. This property of the system is explained by effectively 
reducing the dynamics to the nonlinear problem of a particle moving in a double square well potential. The analysis is made 
for both attractive and repulsive interatomic forces, and it highlights the experimental relevance of our findings.
\end{abstract}

The term Macroscopic Quantum Tunneling (MQT) usually refers to a property of a macroscopic system, or part of it, that performs 
tunneling through a potential barrier like a single particle. After the first experimental discovery of this phenomenon 
\cite{mqtJJfirst} as a tunneling escape from a metastable state in Josephson Junctions (JJ), there have been a number of similar 
realizations in completely different physical systems, such as liquid Helium \cite{helium} and nanomagnets \cite{nano}. 
Frequently MQT is associated with the tunneling between different states of the system with no reference to a spatial energy barrier 
\cite{bec,mqtJJsci,ustinov,baron}, but there are also several examples where macroscopic objects, such as vortices \cite{vortex1,vortex2} 
or fluxons \cite{moon} in JJ's, or magnetic domain walls \cite{magn}, tunnel through a potential barrier.  

Recently \cite{ober,cata}, following earlier theoretical predictions \cite{smerzi}, it has 
been found that a BEC trapped in an harmonic well potential of  mesoscopic length behaves like a single JJ: for nonzero initial imbalance 
of the number of atoms in different wells, Josephson oscillations are present in the system, i.e. the condensate tunnels back and forth 
through the barrier. The only difference with respect to superconducting JJ's is that, for large initial imbalance, the condensate 
is mainly trapped in one of the wells, producing what is called Macroscopic Self-Trapping (MST). 

\begin{figure}[h]
\begin{center} 
\epsfig{file=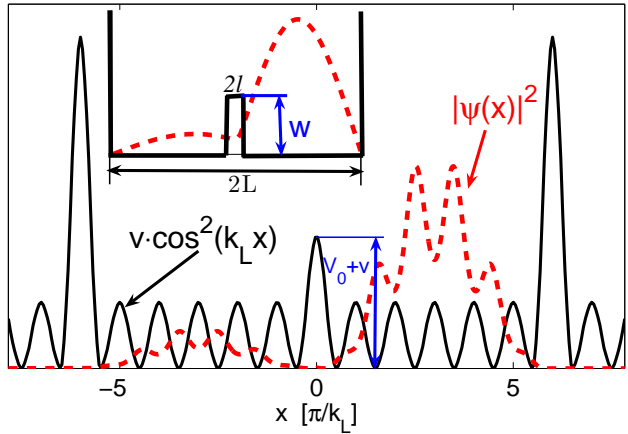,width=0.6\linewidth} 
\end{center} 
\caption{Schematics of the suggested experimental setup. The optical lattice is supplemented by large energy barriers from 
both sides and a small one in the middle (solid curve). The condensate is initially loaded mainly into the right part of the optical lattice 
(the dashed line represents particle density). The inset shows the reduction of the problem 
to the particle motion in a double square well potential (details are given in the text).} 
\label{becexp}\end{figure}

Our aim in this Letter is to suggest the experimental realization of a BEC in an optical lattice, which is engineered in such a way to
mimic two weakly coupled chains of JJ's (see Fig. \ref{becexp}). Using such an experimental set-up, we demonstrate 
the feasibility of the efficient control of a switch between tunneling 
(MQT) and trapped (MST) states of the system. We show that our problem reduces to the Gross-Pitaevskii equation 
(GPE) \cite{gros} in a double square well, which displays very different properties from the previously considered double
harmonic well potential \cite{ober,smerzi,alberto,kevrekidis,min}.
Specifically, we show that for both attractive 
and repulsive nonlinearities the stationary solutions describing the MQT and MST regimes are characterized by very close energies 
in a wide range of the nonlinearity parameter. This property itself allows one to switch from the oscillatory tunneling regime to 
the trapped one and back via a simple pulselike adiabatic change of the energy barrier. Our results are broadly applicable and open the way to
the experimental study of these phenomena in BEC dynamics.

We start from the following, one-dimensional Hamiltonian of a BEC in an optical lattice: 
\begin{equation}
i\hbar\frac{\partial \psi}{\partial t}=-\frac{\hbar^2}{2m}\frac{\partial^2 \psi}{\partial x^2}+V(x)\psi+\frac{2\hbar^2a_s}
{ma_\perp^2}|\psi|^2\psi, \label{1}
\end{equation} 
where $m$ is atomic mass, $a_s$ is the scattering length ($a_s<0$ corresponds to attractive atom-atom interactions and 
$a_s>0$ to repulsive interactions) and $a_\perp=\sqrt{\hbar/m\omega_\perp}$ is the transversal oscillation length,  which implicitly takes
into account the real three dimensionality of the system \cite{equation}, $\omega_\perp$ being the transversal frequency of the trap. 
The optical lattice potential is
\begin{eqnarray}
V(x)=v\cos^2(k_Lx) \qquad \mbox{for} \qquad |k_Lx|>\pi/2 \nonumber \\ V(x)=\bigl(v+V_0\bigr)\cos^2(k_Lx) \qquad \mbox{for} \qquad |k_Lx|<\pi/2, 
\label{111}
\end{eqnarray}
where $k_L$ is the wavenumber of the laser beams that create the optical lattice and $V_0$ is the height of the additional 
spatial energy barrier placed in the middle of the optical lattice. Besides that, Dirichlet boundary conditions with 
$\psi(\pm L)=0$ are chosen in order to describe the large confining barriers at both ends of the BEC. These boundary 
conditions could be realized experimentally by an additional optical lattice with larger amplitude and larger lattice constant, as shown in Fig.\ref{becexp}.   

Introducing a dimensionless length scale $\tilde x=2k_Lx$ and time $\tilde t=E_B t/\hbar$, where $E_B=8E_R=4\hbar^2k_L^2/m$ and $E_R$ is the 
recoil energy \cite{bloch}, we can rewrite \eqref{1} as follows
\begin{equation}
i\frac{\partial \Psi}{\partial \tilde t}=-\frac{1}{2}\frac{\partial^2 \Psi}{\partial \tilde x^2}+\tilde V(\tilde x)\Psi+g|\Psi|^2\Psi, \label{2}
\end{equation}
where the normalized wave-function, $\int |\Psi(\tilde x)|^2d\tilde x=1$, is  introduced \cite{Schlagheck}. 
The dimensionless potential $\tilde V$ still has the form \eqref{111} with the following dimensionless depths of the optical lattice
\begin{equation}
\tilde v=\frac{v}{E_B}, \qquad \tilde V_0=\frac{V_0}{E_B}, \qquad
g=\frac{Na_s}{k_La_\perp^2}\;, 
\label{3}
\end{equation}
$g$ being the dimensionless nonlinearity parameter.

We have performed numerical simulations of Eq. \eqref{2} with 12 wells (6 wells on each side of the barrier as presented in Fig. \ref{becexp}) 
and the parameters $\tilde v=0.25$ (in physical units this means that the depth of the optical lattice is $v=2E_R$), $\tilde V_0=0.15$ 
and we fix the nonlinearity to the value $g=-0.025$, i.e., we choose attractive interactions. The dynamics is similar for repulsive interatomic 
forces (see the discussion below). The phenomenon we study in this Letter does not depend significantly on the actual size of the system, if at least 3 lattice sites are present at each side of the barrier.
\begin{figure}[ht]
\begin{center}
\epsfig{file=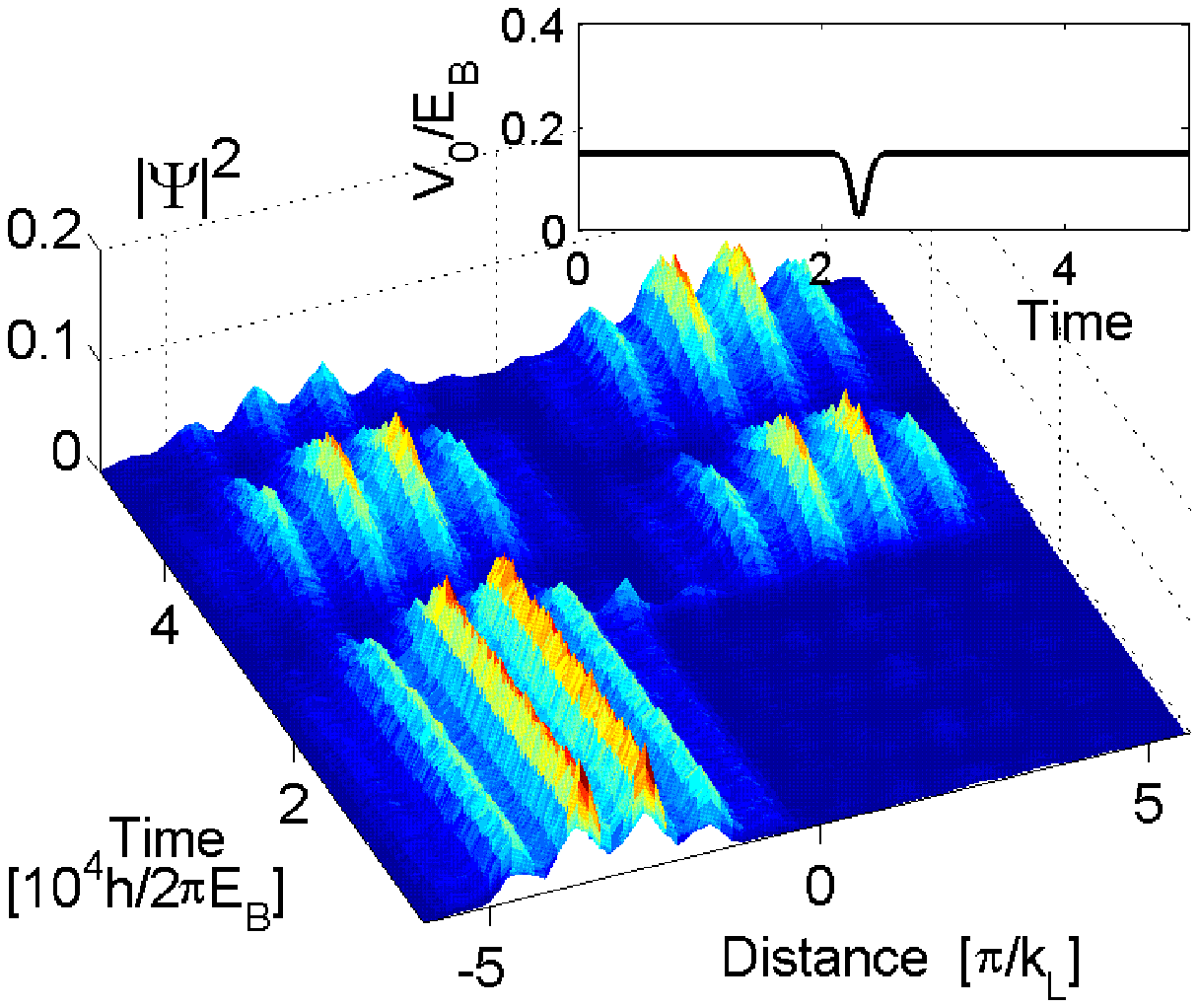,width=0.39\linewidth} \hspace{2cm}
\epsfig{file=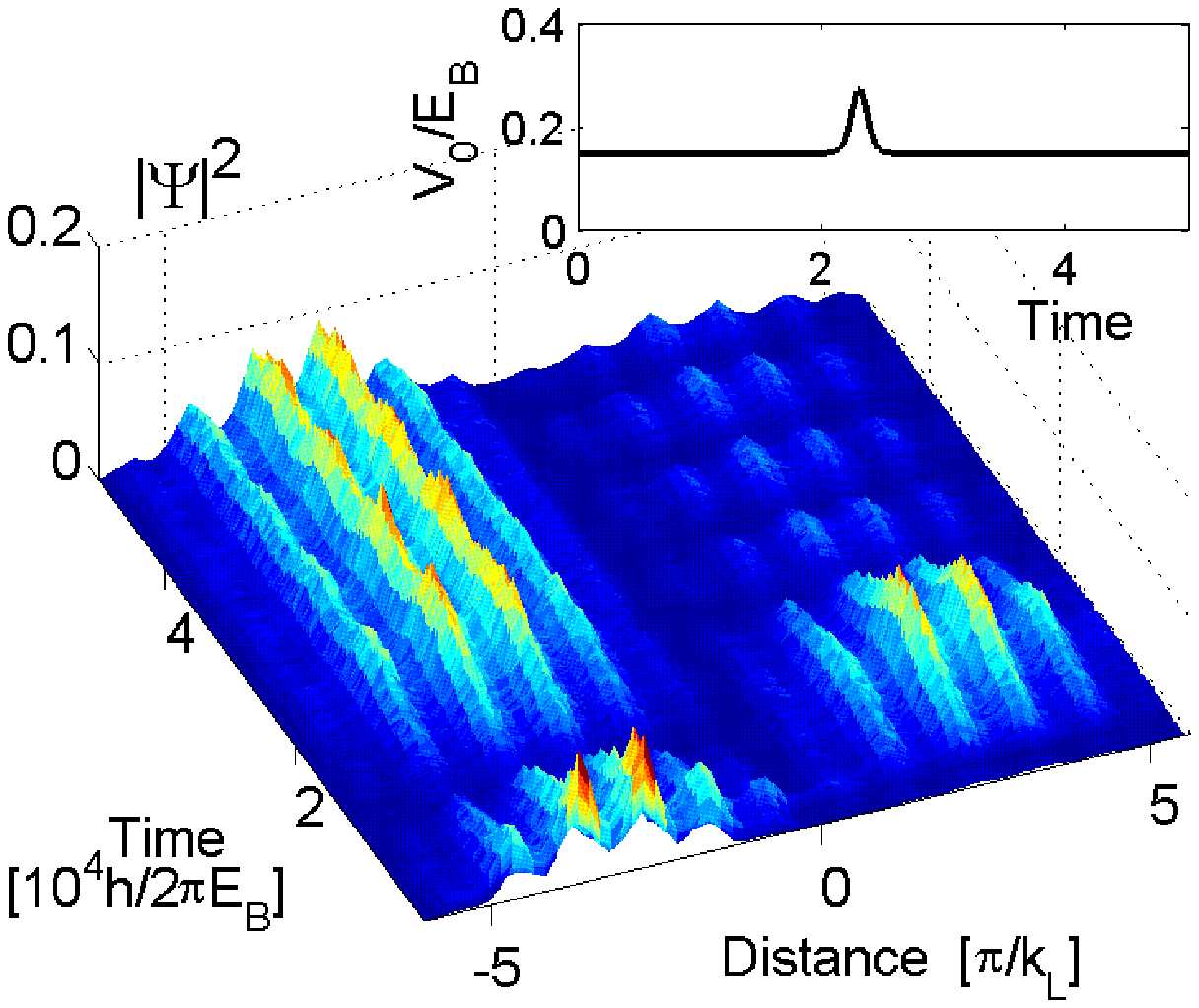,width=0.39\linewidth}
\end{center} 
\caption{(color online) Numerical simulations of Eq. \eqref{2}: the left graph represents the transition from a self-trapped 
state to the macroscopic tunneling regime, while the right graph describes the inverse process. The insets in both graphs 
show the variation of the energy barrier necessary to realize the switching between the different regimes.} 
\label{becosc}\end{figure}

As seen from the left panel of Fig. \ref{becosc}, if one prepares the condensate in a self-trapped state it remains there until we 
apply the pulselike time variation of the barrier displayed in the inset. After that action, the condensate goes into the oscillating 
tunneling regime. On the other hand, preparing the condensate in the oscillating tunneling regime (right graph in Fig. \ref{becosc}) 
one can easily arrive at a self-trapping state by varying again the energy barrier in the middle as displayed in the inset. Let 
us mention that, as far as the energy of the barrier is changed adiabatically, the total energy of the condensate does not vary, 
i.e. the self-trapped and tunneling oscillatory regimes have the {\em same} energy. This is quite different from what happens 
in a double harmonic well potential \cite{ober,smerzi,alberto,kevrekidis,min}. The point is that, in the double harmonic well, 
the asymmetric stationary solution is characterized by a smaller energy than the symmetric solution and this difference 
increases sharply with increasing nonlinearity. Hence, a drastic energy injection is required in order to realize the transition between the 
two regimes; whilst in our case the transition is simply achieved only by varying pulsewise the energy barrier. 
Below we argue that this happens because our case effectively reduces to the case of a double square well potential (see the inset 
of Fig. \ref{becexp} and the reduction procedure below) for which asymmetric and symmetric stationary solutions carry almost 
the same energies in a wide range of the nonlinearity parameter.  

Now we proceed to reducing Eq. \eqref{2} to a Discrete NonLinear Schr\"odinger equation (DNLS). We discretize it via 
a tight-binding approximation \cite{smerzitight,yuri,mark}, representing the wave function $\Psi(\tilde x)$ as
\begin{equation}
\Psi(\tilde x)=\sum_j\phi_j\Phi_j(\tilde x), 
\label{6}
\end{equation}
where $\Phi_j(\tilde x)$ is a normalized isolated wave function in an optical lattice in the fully linear case $g=0$ and could 
be expressed in terms of Wannier functions (see, e.g., \cite{wanier}). For clarity, we use here its approximation for a harmonic 
trap centered at the points $r_j=j\pi(|j|+1/2)/|j|$ ($|j|$ varies from 1 to $n$, the number of wells). In the context of the 
evolution equation \eqref{2} $\Phi_j(\tilde x)$ has the form
\begin{equation}
\Phi_j(\tilde x)=\left(\frac{\sqrt{\tilde v}}{\pi\sqrt{2}}\right)^{1/4}e^{-\sqrt{\tilde v}(\tilde x -r_j)^2/\sqrt{8}}  \;,
\label{7}
\end{equation}
for $|j|\neq 1$, and one should substitute $\tilde v$ by $\tilde v+\tilde V_0$ in the above expression in order to get an 
approximate formula for the wave function for $|j|=1$.
\begin{figure}[t]
\begin{center}
\epsfig{file=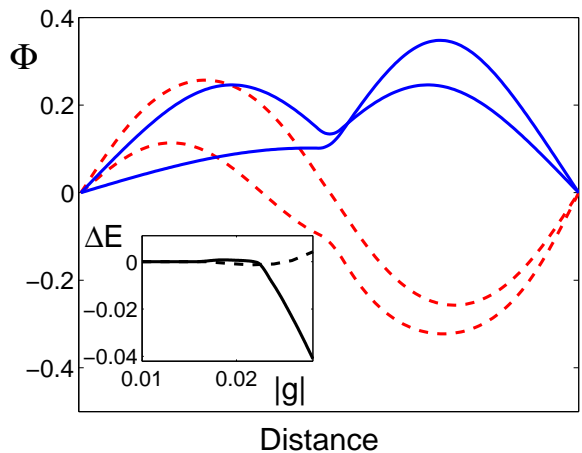,width=0.6\linewidth} 
\end{center} 
\caption{Stationary profiles described by expressions \eqref{attractive} and \eqref{repulsive}. The solid 
line represents symmetric and asymmetric solution profiles for attractive nonlinearities, while the dashed curves 
describe the case with repulsive interactions. The inset shows the relative energy difference between asymmetric 
and symmetric stationary solutions (solid and dashed lines correspond again to attractive and repulsive nonlinearities, respectively).} 
\label{becE}\end{figure}

Assuming further that the overlap of the wave functions in neighboring sites is small, we get from \eqref{2} the following 
DNLS equation for the sites $|j|\neq 1$
\begin{equation}
i\hbar\frac{\partial \phi_j}{\partial \tilde t}=-Q\bigl(\phi_{j+1}+\phi_{j-1}\bigr)+U|\phi_j|^2\phi_j, \label{8}
\end{equation}
while for $|j|=1$ we have
\begin{equation}
i\hbar\frac{\partial \phi_{\pm 1}}{\partial \tilde t}=-Q\phi_{\pm 2}-Q_1\phi_{\mp 1}+U_1|\phi_{\pm 1}|^2\phi_{\pm 1}, \label{81}
\end{equation}
where we assume pinned boundary conditions. The constants $Q$, $Q_1$, $U$ and $U_1$ are easily computed from the following expressions ($|j|\neq 0$):
\begin{eqnarray}
Q&=&-\int\left[\frac{\partial\Phi_j}{\partial \tilde x}\frac{\partial\Phi_{j+1}}{\partial \tilde x}+\tilde v\cos^2(\tilde x/2)\Phi_j\Phi_{j+1}\right]d\tilde x, \nonumber \\
Q_1&=&-\int\left[\frac{\partial\Phi_1}{\partial \tilde x}\frac{\partial\Phi_{-1}}{\partial \tilde x}+(\tilde v+\tilde V_0)\cos^2(\tilde x/2)\Phi_1\Phi_{-1}\right]d\tilde x, \nonumber \\
U&=&g\int\Phi_j^4\,d\tilde x\simeq U_1=g\int\Phi_{\pm 1}^4\,d\tilde x \;.
\label{9} 
\end{eqnarray}
In order to characterize the solutions of Eqs. \eqref{8} and \eqref{81}, we follow the same procedure used in Ref. \cite{jerome}, which
goes through a continuum approximation. Assuming that $\phi_1=\phi_{-1}$ we finally arrive at
\begin{equation}
\frac{i\hbar}{Q}\frac{\partial \phi(j)}{\partial \tilde t}=-\frac{\partial^2 \phi(j)}{\partial j^2}+W(j)\phi(j)+ R|\phi(j)|^2\phi(j), \label{10}
\end{equation}
where now $j$ is a continuous variable, $W(j)$ is a double square well potential with a barrier height $w=2(Q-Q_1)/Q$ and width $l=1$, 
$\phi(j)$ obeys pinned boundary conditions $\phi(j=\pm L)=0$ ($2L$ is a width of a double square well potential) and the nonlinearity parameter 
is given by $R= U/Q$. 

Summarizing, we have reduced the initial problem, GPE with optical lattice and barrier potentials to a DNLS equation and then this latter
again to a  GPE with double square well potential. The reason for doing this, is to get rid of 
the optical lattice potential and to reduce our problem to the GPE with a double square well potential, for which one can easily 
find exact stationary solutions. They are sought as $\phi(\tilde t,j)=\Phi(j)
\exp(-i\beta \tilde t)$ with a real-valued function $\Phi(x)$ found in terms of
Jacobi elliptic functions \cite{book}, in the case of attractive atom-atom interactions $R<0$
\begin{align}
-L<x<-l\ :\ &\Phi=A\ \cn [\gamma_A(j+L)-{\mathbb K}(k_A),k_A],  \nonumber\\
l<x<L\ :\ &\Phi=B\ \cn [\gamma_B(j-L)+{\mathbb K}(k_B),k_B] ,\nonumber \\
-l<x<l\ :\ &\Phi=C\ \dn [\gamma_C(j-j_0),k_C],
\label{attractive}
\end{align}
with the parameters given in terms of the amplitudes by
\begin{eqnarray}
\gamma_A^2=\beta+\frac{A^2}{|R|}, \quad k_A^2&=&\frac{A^2}{2(A^2+|R|\beta)}, \qquad \qquad
\gamma_B^2=\beta+\frac{B^2}{|R|}, \quad k_B^2=\frac{B^2}{2(B^2+|R|\beta)}, \nonumber \\
\gamma_C^2&=&w-\beta-\frac{C^2}{2|R|}, \quad k_C^2=\frac{w-\beta-C^2/|R|}{w-\beta-C^2/2|R|}, \nonumber
\end{eqnarray}
while, in case of repulsive interactions $R>0$, one has the stationary solution written in the form
\begin{align}
-L<x<-l\ :\ &\Phi=A\ \sn [\gamma_A(j+L),k_A],  \nonumber\\
l<x<L\ :\ &\Phi=B\ \sn [\gamma_B(j-L),k_B] ,\nonumber \\
-l<x<l\ :\ &\Phi=C/\cn [\gamma_C(j-j_0),k_C],
\label{repulsive}
\end{align}
where
\begin{eqnarray}
\gamma_A^2=\beta-\frac{A^2}{2|R|}, \quad k_A^2&=&\frac{A^2}{2|R|\beta-A^2}, \qquad \qquad
\gamma_B^2=\beta-\frac{B^2}{2|R|}, \quad k_B^2=\frac{B^2}{2|R|\beta-B^2}, \nonumber \\
\gamma_C^2&=&2\left(w-\beta+\frac{C^2}{|R|}\right), \quad k_C^2=\frac{w-\beta+C^2/2|R|}{w-\beta+C^2/|R|}. \nonumber
\end{eqnarray}
Here ${\mathbb K}$  denotes the complete elliptic integral of the first kind, and, by construction, the above expressions 
verify the vanishing boundary values in $j=\pm L$. 

The solutions are then given in terms of five parameters ($A$, $B$, $C$,
$\beta$, $j_0$), which are determined by the four continuity conditions in
$j=\pm l$ and the wave function normalization condition $\int dj\Phi^2(j)=1$. In both the repulsive and attractive case, one has symmetric 
and antisymmetric solutions, and the asymmetric solution appears above a given nonlinearity threshold value. In Fig. \ref{becE}, we plot 
the profiles of the lowest energy symmetric (attractive case) and antisymmetric (repulsive case) solution, the asymmetric solutions 
for both the repulsive and the attractive cases, and the relative energy differences $\Delta E=2(E_a-E_s)/(E_a+E_s)$ between asymmetric 
$E_a$ and symmetric $E_s$ cases. As seen from the inset, these energies are very close in the nonlinearity range $0.018<|g|<0.03$ (note 
that the numerical simulations presented in Fig. \ref{becosc} are made for $g=-0.025$) and hence it is easy to switch from the tunneling 
regime to the self-trapped state and back again.   

In order to get an idea about a possible experimental realization of these effects, we choose $^7$Li, which is characterized by an
attractive atom-atom interaction \cite{lithium}. With the realistic experimental parameters $\omega_\perp=2\pi\times 30$Hz, $a_s=-1.4$nm, 
$k_L=7.4\cdot\mu$m$^{-1}$, from the formula for nonlinearity $g=Na_s/a_\perp^2k_L$ one gets that the total atom number needed 
to access values of $|g|$ around 0.025 is $N\approx 10000$. While in case of $^{23}$Na \cite{natrium} with repulsive forces ($a_s=4.9$nm), 
the atom number should be $N\approx 1000$. Increasing the number of wells (system size) $n$ times, one should decrease 
the number of atoms $n^2$ times in order to observe the predicted effect.  

To summarize, we predict an effective switching scenario from an oscillatory tunneling to a self-trapped regime as a novel 
macroscopic quantum transport effect to be realized with ultracold atoms. We have shown that the problem effectively 
reduces to the particle motion in a double square well potential in contrast to earlier studies dealing with double harmonic wells. 
This difference guarantees the possibility of the switch from an oscillatory tunneling to a self-trapped state via a pulse-like change of 
the central potential barrier. We have derived typical ranges of physical parameters, for both attractive and repulsive interatomic forces, in order 
to suggest a ready-to-implement experimental verification. 

\acknowledgments
It is our pleasure to thank Oliver Morsch and Ennio Arimondo 
for useful discussions. S.R. acknowledges financial support by the PRIN05-MIUR grant on {\it Dynamics and thermodynamics of systems with long-range interactions}, 
R.Kh. by the Marie-Curie international incoming fellowship award (MIF1-CT-2005-021328) and NATO (grant no. FEL.RIG.980767), and S.W. by the Alexander von Humboldt foundation (Feodor-Lynen Program).

\end{document}